\documentclass[aps,prl,twocolumn,nofootinbib,superscriptaddress,floatfix]{revtex4-1}  
\usepackage{graphicx}  
\usepackage{amsmath}
\usepackage{amssymb}   
\usepackage{color}
\usepackage{hyperref}
\usepackage{ulem}

\def\gsim{\;\rlap{\lower 2.5pt \hbox{$\sim$}}\raise 1.5pt\hbox{$>$}\;}
\def\lsim{\;\rlap{\lower 2.5pt \hbox{$\sim$}}\raise 1.5pt\hbox{$<$}\;}

\begin{document}

\title{On the role of  GRBs on life extinction in the Universe}

\author{Tsvi Piran}
  \email{tsvi.piran@huji.ac.il}
 \affiliation{Racah Institute of Physics, The Hebrew University, Jerusalem 91904, Israel} 

\author{Raul Jimenez}
  \email{raul.jimenez@icc.ub.edu}
 \affiliation{ICREA \& ICC, University of Barcelona, Marti i Franques 1, Barcelona 08028, Spain.} 
 \affiliation{Institute for Applied Computational Science, Harvard University, MA 02138, USA.}

\date{\today}

\begin{abstract}
As a copious source of gamma-rays, a nearby Galactic Gamma-Ray Burst (GRB) can be a threat to life. 
Using recent determinations of the rate of GRBs, their luminosity function and properties of their host galaxies, we estimate
the probability that a life-threatening (lethal) GRB would take place. Amongst the different kinds of GRBs,   long ones are most dangerous. 
There is a very good chance (but no certainty) that   at least one lethal GRB  took place during the past 5 Gyr close enough to Earth as to significantly damage life. There is a 50\% chance that such a lethal GRB took place during the last 500 Myr causing one of the major mass extinction events. Assuming that a similar level of radiation would be lethal to life on other exoplanets hosting life, we explore the potential effects  of GRBs to life elsewhere in the Galaxy and the Universe. We find that the   probability  of a lethal GRB  is   much larger in the inner Milky Way (95\%  within  a radius of 4 kpc  from the galactic center), making it inhospitable to life.
Only at the outskirts of the Milky Way, at more than 10 kpc from the galactic center, this probability  drops below 50\%. 
When considering the Universe as a whole, the safest environments for life (similar to the one on Earth) are the lowest density regions in the outskirts of large galaxies and life  can exist in only $\approx 10\%$ of  galaxies.
Remarkably, a cosmological constant is essential for such systems to exist. Furthermore,  because of both the higher GRB rate and galaxies being smaller, life as it exists on Earth could not take place at $z > 0.5$. Early life forms  must have been much more resilient to radiation.
\end{abstract}

\maketitle

\section{Introduction}

Gamma Ray bursts (GRBs), short and intense bursts of $\gamma$-rays, are the brightest explosions known. The copious flux of $\gamma$-ray photons with energies above $100 keV$ from  a galactic GRB could  destroy the ozone layer making them  potentially damaging to life on Earth. This has led to the suggestion\footnote{See \citep{Ruderman74} for an earlier discussion of   nearby Supernovae as the cause of  life extinction.}   \citep{Thorsett95,Dar+98,Dar01,Scalo+02,Melott04,Thomas05a,Thomas05b,MT09,Karam}   
that events of massive life extinction were caused by galactic GRBs. This issue depends of course on the rate of galactic GRBs in the Earth neighborhood. Once it was realized that long GRBs are preferentially located  at low-metallicity environments it was  claimed \citep{Stanek06}  that  nearby Galactic GRB are rare and  GRBs  are unlikely to play any role in life extinction on Earth (see however, \citep{Melott06} who claims that metallicity won't protect life on Earth from GRBs). 
 Given the  recent significant progress in quantifying 
the  main ingredients that determine whether  GRBs have any effect on Earth: their  rate,  luminosity function and  dependence on metallicity it is therefore timely to re-asses this issue, extending the discussion to GRBs effects on life in the whole Milky Way  and in the whole Universe. 

GRBs are traditionally divided in two  groups according to their duration: long  ($>2$s) GRBs (LGRBs) and short ($<2$s) GRBs (sGRBs).  This division follows to a large extent\footnote{We note in passing that some GRBs that are shorter than $2$s do arise from collapsing massive stars \cite{Bromberg+13}. However this is unimportant for this work. } the origin of these events. LGRBs are associated with the death of massive stars \citep[see e.g.][for a review]{WoosleyBloom} while sGRBs have a different origin, most likely compact binary mergers \citep{Eichler89}.  Recently, it was realized that there is 
a third group characterized by  low luminosity ($L\approx 10^{46-48}$ erg s$^{-1}$) and denoted  {\it 
ll}GRBs. These events are also associated with the death of massive stars, but they originate from a different  physical mechanism \citep{Bromberg+11}. 
A  fourth type of a related explosion - giant SGR flares might  also relevant. Such a flare took place in the Milky Way on  27 Dec 2004,  releasing $\approx 4 \times 10^{46}$ ergs \citep{Palmer+05}. This flare, that was sufficiently powerful to disturb the Earth ionosphere, seen as a brief change in the ionization levels in the lowest regions of the Earth's ionosphere (the D-layer), is the only known object outside the solar system to have a direct clear impact on Earth.  In fact, this type of disturbance was first seen from a GRB830801 in 1983  \cite{FI88}. Giant SGR flares are a different phenomenon than GRBs but as their  rates  could be as high as once per thirty years in the Galaxy, we explore their possible role as well. Solar flares are another potential life threatening source as they are stronger than previously thought \cite{MT11,MT12,Usoskin}.

Wanderman and Piran \cite{WandermanPiran} have recently reconstructed, in a  model independent way, the rate of LGRBs as a function of redshift and their luminosity function. One of their most interesting findings is that the LGRB rate is not reproduced by the star formation rate of the global galaxy population. This discrepancy is statistically highly significant,  particularly  at low ($<3$) redshifts, which is relevant here. This is, at first,  surprising as there is ample evidence that long duration GRBs originate from the collapse of very massive stars and  
one would expect that LGRB  follow  the SFR. 
\citet{jp} have shown that the LGRB rate and the galaxy derived star formation rate (SFR) agree for a  special class of galaxies: low mass (stellar mass $< 10^{10}$ M$_{\odot}$) and low metallicity ($\lsim 1/10$ solar). This is, of course, done in a statistical sense and does not exclude that few outliers to this trend exist. But it is clear that the LGRB host population is a special subclass of the general galaxy population.   
These results are in agreement with earlier observations that indicate that LGRBs take place in dwarf \citep{Natarajan+97}, 
low metallicity  \citep{Fynbo+03} galaxies. They are also consistent with direct observations of LGRBs host metallicities \citep[e.g.][]{savaglio,Leves,Cucchiara} and with  the findings of \citet{Fruchter06} who have shown that the local SFR in the vicinity of LGRBs is much higher than expected if they simply follow the general SFR of the host galaxy \citep[see also][]{Svensson10}.

sGRBs have very different host environments and they clearly 
arise from different progenitors \citep[see e.g.][for reviews]{Nakar07,Berger13}. They are significantly weaker than LGRBs and as such are observed to much shorter distances than LGRB. sGRBs are believed to originate in compact binary mergers \citep{Eichler89} but a direct proof for that is still lacking. As sGRBs are weaker, fewer GRBs have been observed than LGRBs. However their current overall rate is about five times larger than the rate of LGRBs.   In the following we use a recent determination of the sGRBs global rate and luminosity function by \citet{WP14}.

{\it ll}GRBs are significantly weaker with energies of  $10^{47-49}$erg  (as well as smoother and softer) than both LGRBs and sGRBs. Like LGRBs they are associated with the death of massive stars but they  arise due to a different physical mechanism \citep{Bromberg+11}. While less than half a dozen {\it ll}GRBs have been observed so far they  are more numerous than both  LGRBs or sGRBs \citep{Soderberg+06}. Because of their low luminosities they  are observed only up to relatively short (but still cosmological) distances. 

We use the very recent determination of GRB rates and luminosity function to estimate the flux of Galactic GRBs on Earth and compare it with the flux
needed to destroy the ozone layer. Given that LGRBs are the most powerful and hence most dangerous, and given their dependence on metallicity we begin with an exposition of the Milky Way metallicity distribution. 
We continue estimating the life threatening effect of LGRBs, turning later, using the same formalism  to sGRBs,  {\it ll}GRBs and giant SGR flares. We conclude summarizing the results and their implication  to life extinction on Earth. We also explore the implications   to life extinction on exoplanets elsewhere in the Milky Way  and in the whole Universe.

\section{The Milky Way Metallicity Distribution}

LGRB rate estimates derive the expected rates of LGRBs per unit volume per unit time. When translating this volumetric rates to event rate per galaxy and more specifically to the rate within the Milky Way,  one has to consider the type of galaxies in which the events take place. 
Our earlier analysis \citep{jp} shows that LGRB hosts are dwarf low metallicity galaxies that are very different from the Milky Way. 
There are outliers and some LGRBs has been found in higher metallicity galaxies \citep{savaglio,Leves}. 

\begin{figure}
\includegraphics[width=1.05\columnwidth]{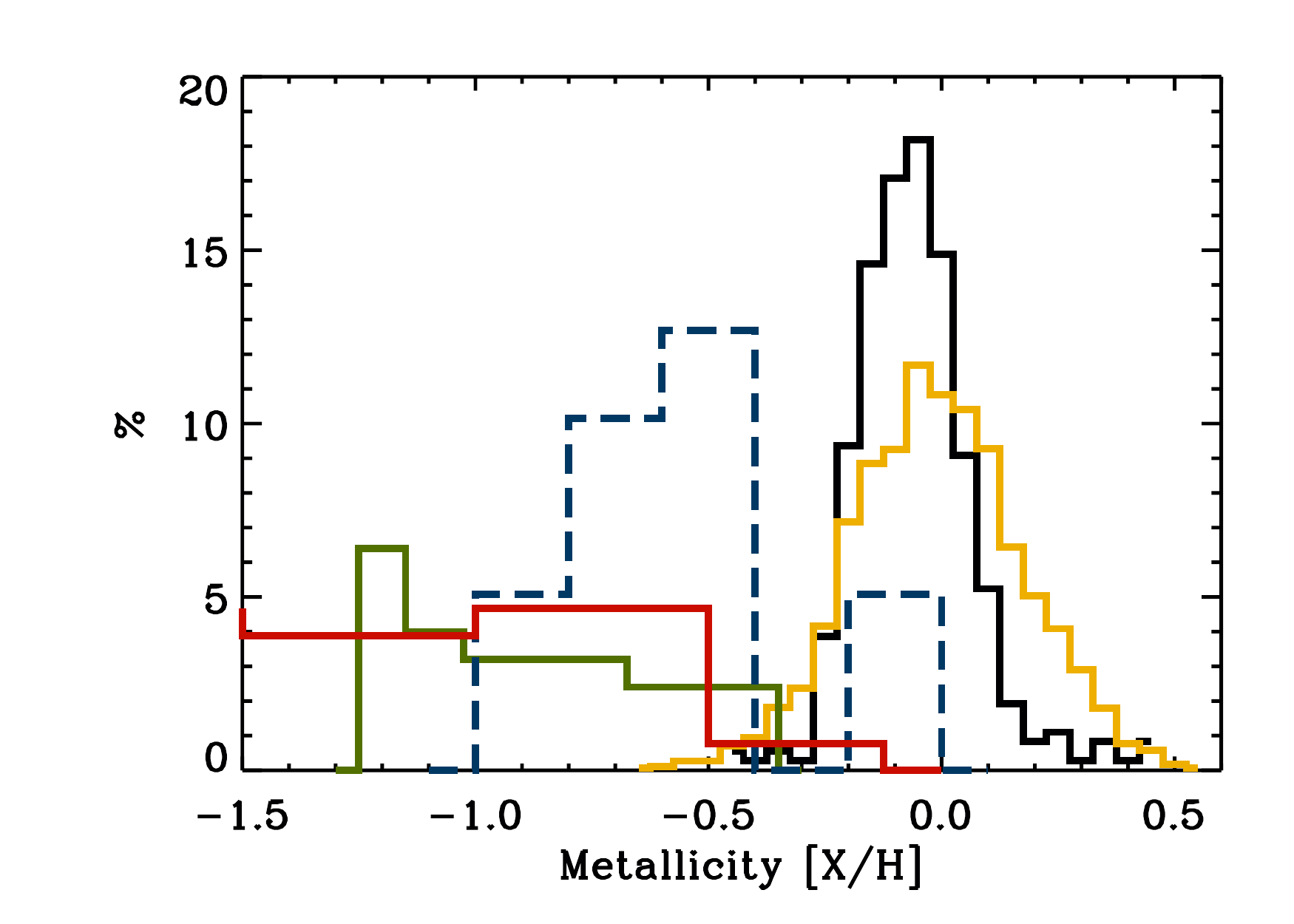}
\caption{\label{fig:mw} The percentage of stars as a function of metallicity in the Milky Way disk with ages $1 <$ t/Gyr $ < 5$ (solid orange line) and with ages $ < 1$ Gyr (solid black line) as obtained by \citet{luca}. The distribution of LGRBs metallicity as obtained by \citet{jp} from matching the RGB global rate to the global star formation rate of galaxies (solid green line) and that from direct metallicity determinations of LGRBs (dashed line) \citep{savaglio} and \citet{Cucchiara} from Damped Ly-$\alpha$ systems (DLA) (solid red line). The overlap between the LGRB and Milky Way stars distributions is only at the few \% level.}
\end{figure}

Ref.~\cite[][and references therein]{luca}   determine the ages and metallicities of stars in the Milky Way disk.  Fig.~\ref{fig:mw} depicts the percentage distribution of stars in the Milky Way for ages $< 1$ Gyr (solid black line) and stars older than 1 Gyr but younger than  5 Gyr (solid orange line). Stars that are older than the Sun and that therefore trace the chemical conditions of the star forming gas at earlier epochs are not relevant for the question of life destruction on Earth. In the same plot we also show (solid green line) the percentage distribution of LGRB hosts derived from \cite{jp} using the mass metallicity  relation from \cite{panter}. Note that due to the metallicity bias for the LGRB host galaxies, there is very little overlap with the distribution of stars in the Milky Way disk. In fact they only overlap at the 10\% level. 

Also shown in Fig.~\ref{fig:mw} are the distributions of LGRB hosts with direct metallicity determinations (dashed blue lines) as compiled by \cite{savaglio} and those of GRB hosts metallicities derived from DLA measurements (red line) as reported by \cite{Cucchiara}. 
 The percentage of overlap of direct hosts metallicities with those of stars in the Milky Way is 10\%.
We conclude that  the metallicity bias will reduce the probability for LGRB within the last 5 Gyrs in the Milky Way by a percentage between 5\%  (from the metallicity determination by \cite{jp}) and 10\%  (from direct metallicity determinations 
\citep[e.g.][]{savaglio,Leves,Cucchiara}), resulting in a reduction factor between  $10$ and $20$ as compared to the volumetric rate of LGRBs. 
In what follows we will assume a conservative 10\% value for a metallicity bias for LGRB above solar.

\section{Life threatening GRBs in the Milky Way}

\begin{figure}
\includegraphics[width=\columnwidth]{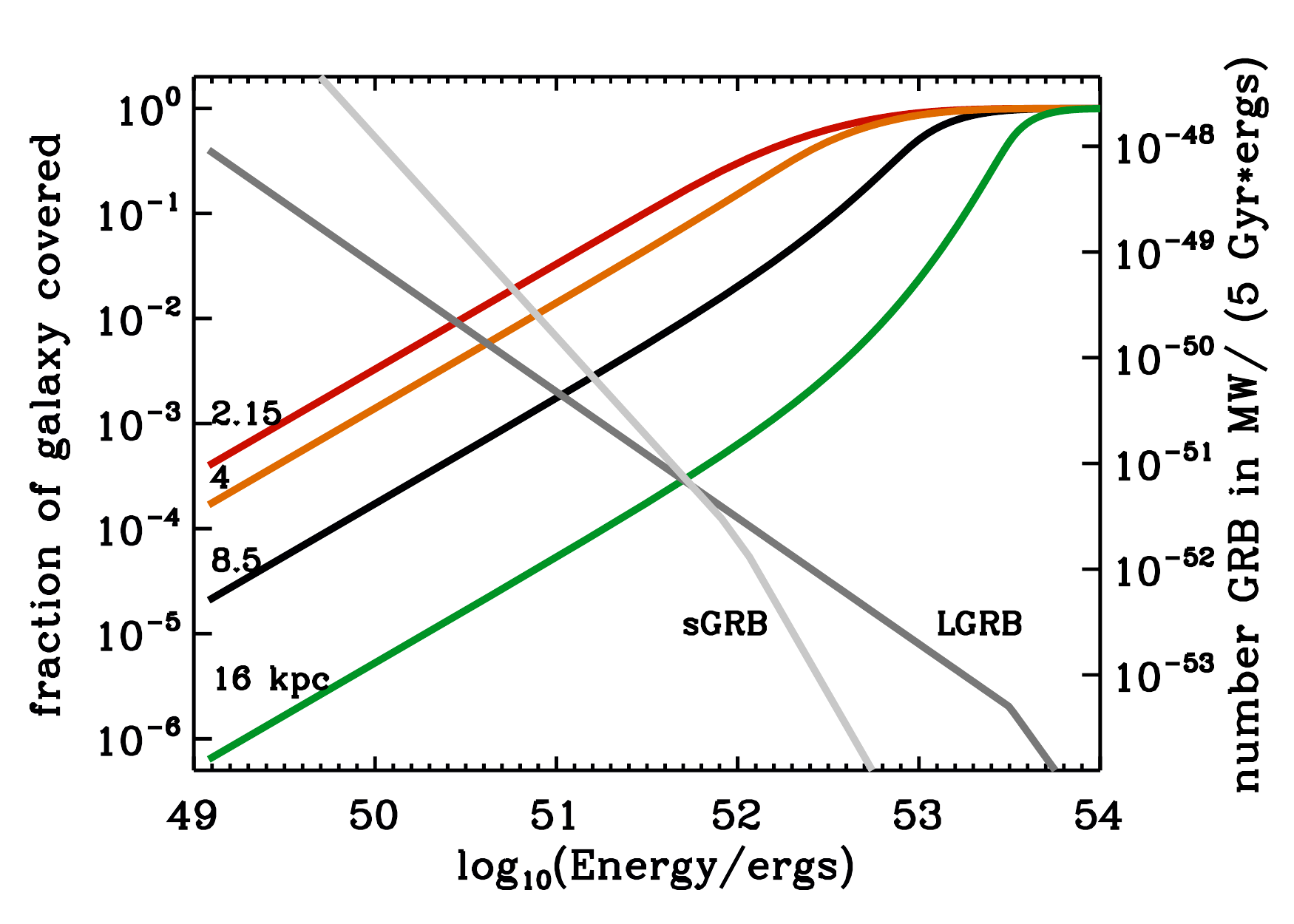}
\caption{\label{fig:num} Left y-axis: the mass fraction of the galaxy from which the fluence
on a planet will exceed  $100$ kJ/m$^2$ for a given explosion energy (x-axis). The colored curves correspond to different locations of the life harboring exoplanet  ($2.15, 4, 8.5$ and $16$ kpc from the Galactic center). We have adopted for the MW an exponential disc with scale-length of $2.15$ kpc. The right y-axis provides (for the gray curves) the number of GRBs in the MW in the past $5$ Gyr per erg. For a given energy, the product of the corresponding colored and gray curves gives the number of damaging GRBs to life per energy interval.} 
\label{fig:distribution}
\end{figure}

Following \citet{WandermanPiran,WP14} we write the current ($z=0$) luminosity function as: 
\begin{equation}
\phi(L) = n_0 
\left\{
\begin{array}{ll}
({L}/{L^*})^{-\hat \alpha}   &  L_{\rm min}< L < L^*  \\
({L}/{L^*})^{-\hat \beta}     &  L^* < L < L_{\rm max}  .
\end{array}
\right.
\label{eq:lf}
\end{equation}
The parameters of the luminosity functions\footnote{ The luminosity function defined here, $\phi(L)$,  is per $dL/L*$.  As such it differs from that given in \citep{WandermanPiran,WP14} that is per $d\log_{10} (L)$. The power law indices are marked by $\hat {}$ to denote this difference. Clearly,  $\hat \alpha = \alpha +1$ and $\hat \beta = \beta +1$.} are given in Table~\ref{table_lumF}  and the functions are  shown in Fig.~\ref{fig:distribution}. 
This luminosity and rate are the isotropic equivalent (namely disregarding the poorly constrained beaming), which are the quantities needed for our estimates here. 
In the following we need the total energy (see also Ref.~\cite{Ejzak}) and not the peak luminosity. A good but rough estimate is obtained by assuming a typical duration of $20$s  ($1$s) for LGRBs (sGRBs). Multiplying by the average ($\sim$ half) of the peak flux we obtain  $E_{LGRB}= 10 L$ and  $E_{sGRB} = 0.5 L$.
In what follows we  adopt the cosmological volume occupied by a Milky Way type galaxy as $10^{-7}$ Gpc$^3$ (see e.g. \citet{moped} Fig. ~3 where we use  $6 \times 10^{10}$ M$_{\odot}$ as the  stellar mass of the Milky Way   \citep{MW}).

\begin{table*}
\caption{Parameters of the LGRBs and sGRBs luminosity functions from \citet{WandermanPiran,WP14}. Note that the upper and lower limits are not well determined but this is unimportant for our estimates here. }
\begin{center}
\begin{tabular}{ccccccc}
\hline 
\hline
	& $n_0$ 					&$ \hat \alpha $ & $ \hat \beta$ & $L^* $ 		   & $L_{\rm min}$   & $L_{\rm max}$  \\
	& {\rm Gpc}$^{-3} {\rm yr}^{-1}$	&                        & 			&  ergs s$^{-1}$            & ergs s$^{-1}$           & ergs s$^{-1}$    \\
\hline
LGRB &   $ 0.15^{+0.7}_{-0.8}$   				&   $1.2^{+0.2}_{-0.1}$   	          &   $2.4^{+0.3}_{-0.6} $  		&   $10^{52.5\pm 0.2}$    &   $10^{49}$        &   $10^{54}$       \\
sGRB &    $0.04^{+0.023}_{-0.019}$   				&   $1.9\pm0.12$  		 &   $3.0^{+1}_{-0.8}$  		&   $10^{52.3\pm0.2}$    &   $5 \times10^{49}$    &   $10^{53}$ \\
\hline
\end{tabular}
\end{center}
\label{table_lumF}
\end{table*}

Assuming that GRBs follow the stellar distribution, they Êare distributed in the exponential disk of the Milky Way with a radial density profile given by $\rho \propto \exp (-r/r_d)$, with $r_d = 2.15 \pm 0.14$ kpc (a number that, surprisingly, has only been accurately determined recently \citep{Rix}). 
Using this density profile we calculate  $p[d,R]$,  the fraction of the Galaxy within a distance $d$ from a position $R$ (see Fig.~\ref{fig:num}). The expected number of 
GRBs, with a fluence  exceeding ${\cal F}$  at a location at distance $R$ from the Galactic center 
is: 
\begin{equation} 
\langle  N \rangle = \int_{L_{min}}^{L_{max}} \phi(L) ~ p[d(E,{\cal F}),R] ~dL .
\label{eq:N}
\end{equation}

To estimate the effect of a GRB on life on Earth we need to know what the dangerous radiation doses are. \citet{Ruderman74}, who considered at the time the effect of a nearby SNe on Earth, realized that the most damaging effect would be 
the depletion of the Earth protective Ozone layer for a period of months. This would happen via formation of stratospheric nitric oxide that destroys the Ozone. The Ozone depletion would lead to enhancement of UVB solar radiation that, in turn, would be harmful to life.  Note that the UVB fluence on the surface of the ocean will destroy surface marine life \citep[as described in detail in Ref.][]{Thomas05b} among them plankton, which will deprive (marine) life of their main nutrient. In 1995, after it was realized that GRBs are cosmological and their rate was estimated, \citet{Thorsett95} applied these ideas to 
Galactic GRBs. A decade later \citet{Thomas05a,Thomas05b} carried out the most extensive, to date, calculation of the effects of the gamma-ray flux on the Earth atmosphere. They find that a fluence of $10 kJ/ m^2$ will cause a depletion of -68\% of the ozone layer on a time scale of a month. Fluences of $100 kJ/ m^2$ and $1000 kJ/ m^2$  will cause  depletions of -91\% and -98\% respectively. One has to realize that these are average quantities. 
The exact amount of depletion depends on the direction of the GRB as well as on the season when the GRB takes place and may vary from one latitude to another. 
Following \citet{Thomas05a,Thomas05b} we estimate that  a fluence of  $10 kJ/ m^2$ will cause some damage to life, while $1000 kJ/ m^2$ will wipe out nearly the whole atmosphere causing a catastrophic life extinction event; we consider  ${\cal F} =  100 kJ/ m^2$  as  our canonical life threatening fluence. We don't consider here other sources of damage, such as the possibility that cosmic rays (CR) are associated with the GRBs and those could lead to enhanced radioactivity in the atmosphere \citep{Dar+98,Dar01}. The mean free path for deflection in the galactic magnetic field for a $100$ GeV proton is $1$ kpc. So the lowest part of the CR spectrum which contains the largest number of CRs will be deflected and won't reach Earth if the event is more than $1$ kpc away. This also means that while we will get eventually CR flux from GRBs that  don't point towards Earth, a single event will always be less powerful (because of deflection away of CRs) so their effect will be weakened and depending on their spectrum significantly weakened.

Integrating over the luminosity functions in eq.~\ref{eq:N}  we  estimate  $\langle  N \rangle $,   for both long and short GRBs. These values are listed in  Table.~\ref{table_results}.  To estimate the significance of these numbers taking into account the errors in the luminosity function, burst duration and the Milky Way disk scale length,  we carry out a Monte Carlo simulation of 1000 realizations  for both long and short GRBs. We calculate the distribution of  $\langle  N \rangle$ and the overall probability of  more than one life threatening GRB taking place within the last 5 Gyr, 1 Gyr  and  500Myr.  

\begin{figure}
\includegraphics[width=\columnwidth]{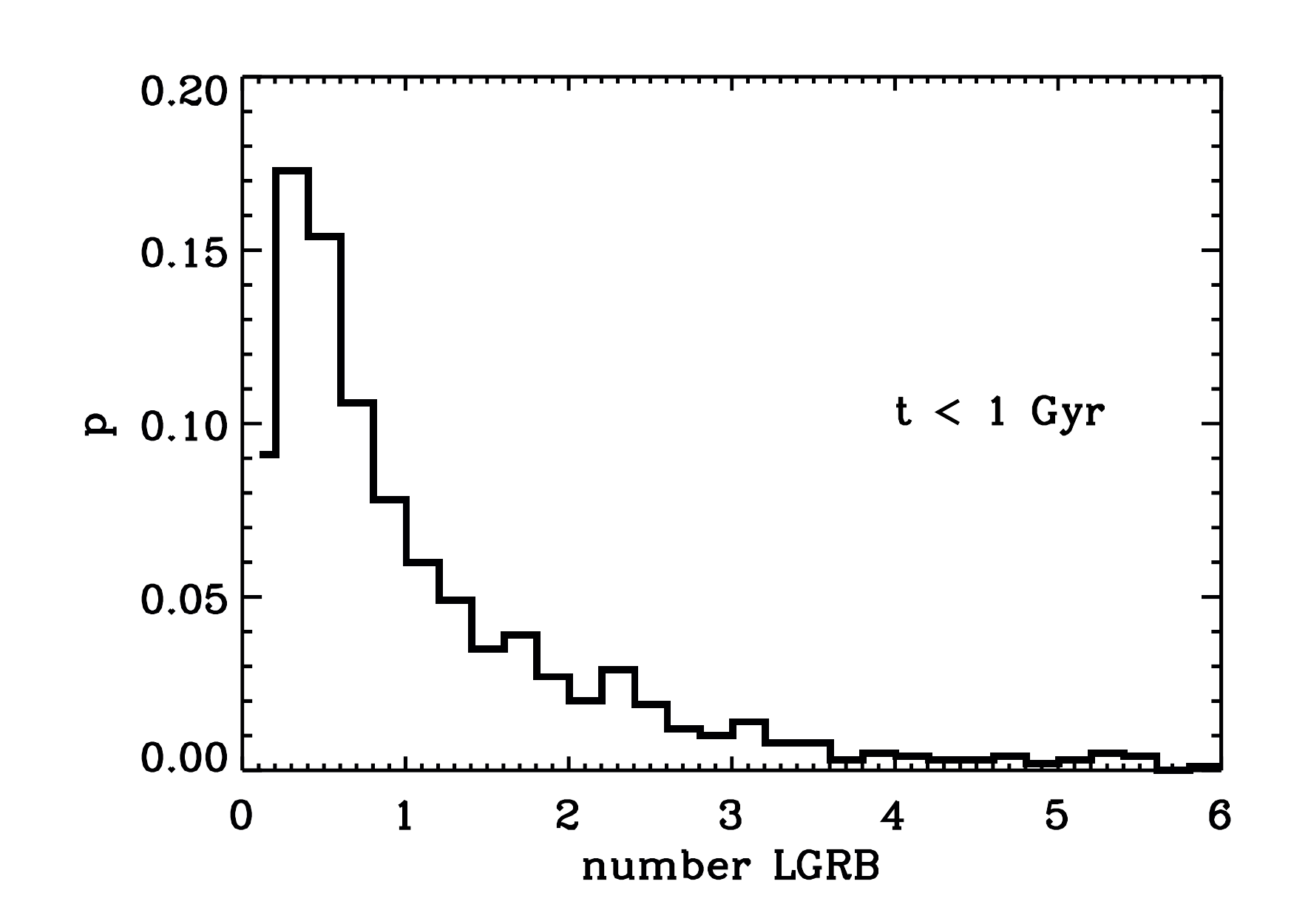}
\includegraphics[width=\columnwidth]{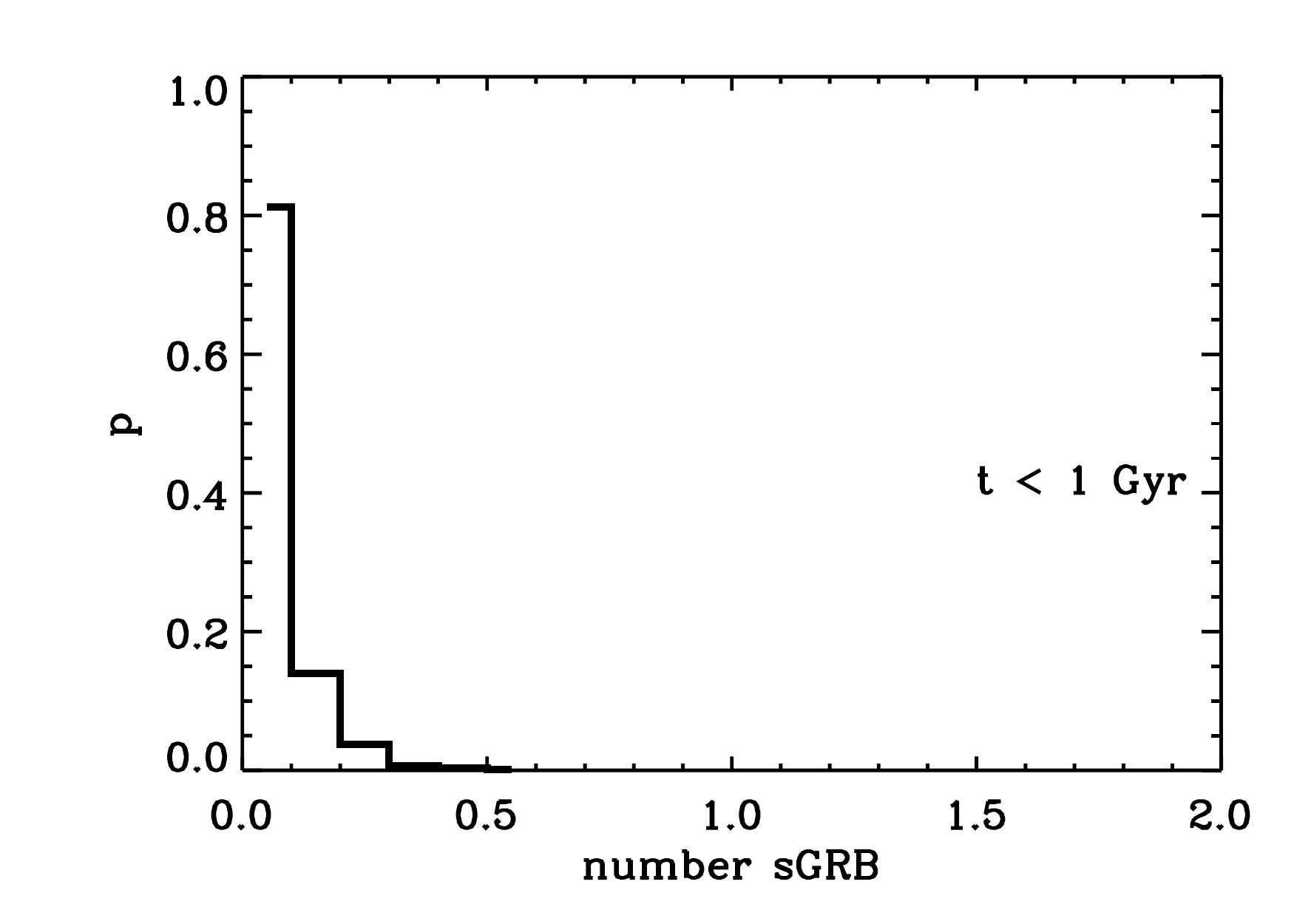}
\caption{\label{fig:LF} The probability distribution function, $p$,  of the average number of lethal LGRBs (top panel) and sGRBs (bottom panel) that irradiated Earth in the past Gyr with enough flux to cause severe life extinction (100 kJ/m$^2$). For LGRBs we show the case where we applied a 10\% metallicity bias.}
\end{figure}

\begin{figure} 
\includegraphics[width=\columnwidth]{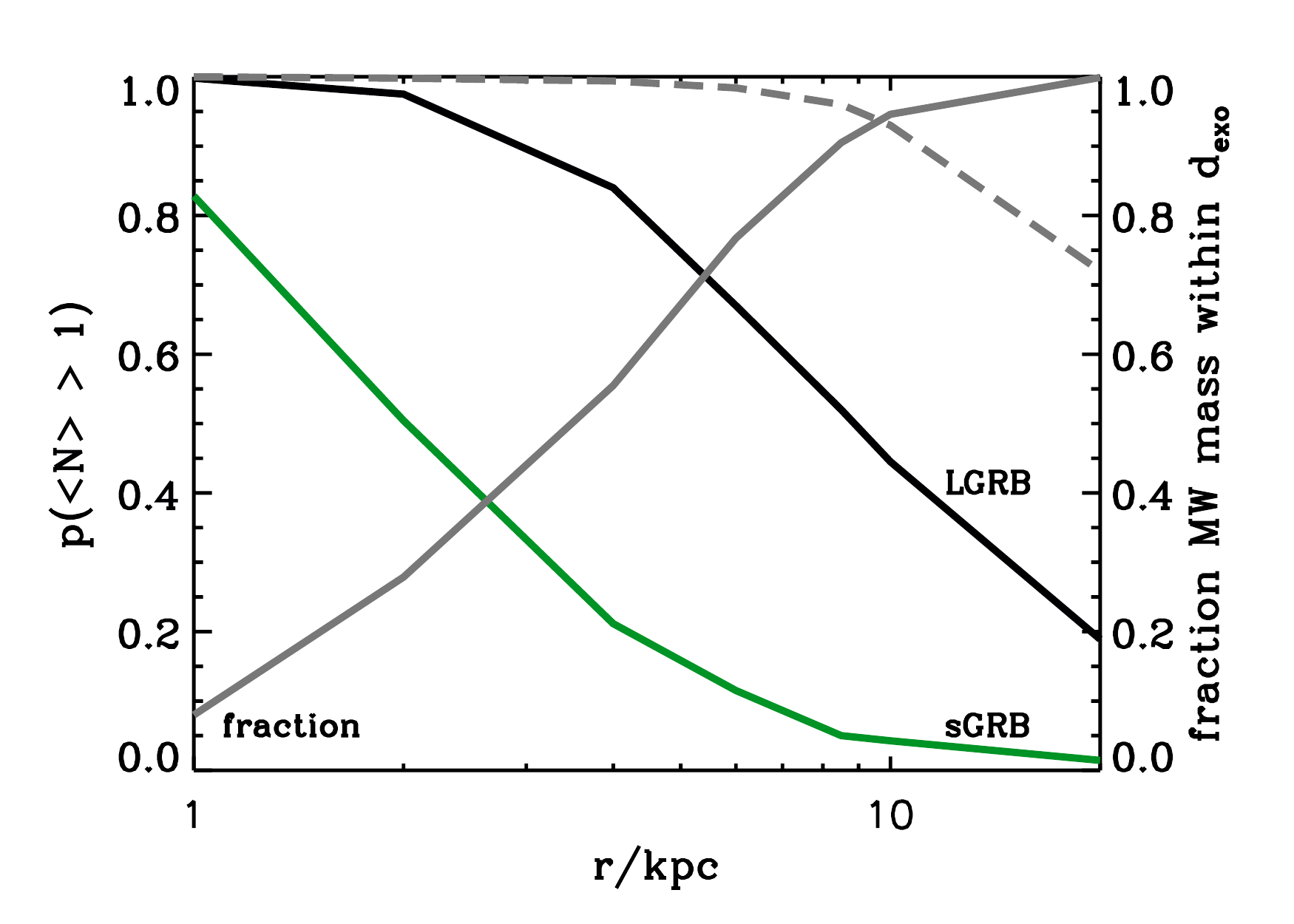}
\caption{\label{fig:MWtot} The probability, $P( \langle N \rangle)$,  of having on  average  more than one lethal GRB in the past Gyr for an exoplanet at a distance $r$ from the centre of the Milky Way. The grey line shows the fraction of mass in the Milky  encompassed within a radius $r$. The dashed line is for LGRB assuming no metallicity correction.}
\end{figure}

Inspection of Fig. \ref{fig:num} reveals that maximal danger arises from $\sim L^*$ bursts. Lower luminosity bursts are more abundant but their covering fraction of the Galaxy is too small. Higher luminosity bursts can destroy life on a large fraction of the Galaxy but those are extremely rare.  From the point of view of computational certainty these results are reassuring as the confidence in  our determination of the rate of events around $L^*$ is good.   This is also important from another point of view. Spatially GRBs are concentrated within regions of the highest SFR \citep{Fruchter06,Svensson10}. The dominance of  strong GRBs whose radius of influence is a few kpc  implies that we can ignore
this spatial inhomogeneity and the approximation that the distribution of LGRBs follow the distribution of matter in the galaxy holds. 

We find that the probability of a LGRB, in the past 5 Gyr, with fluence 
$100 $kJ/m$^2$ on Earth to be higher than 90\% and  in the last 0.5 Gyr this probability is 50\%. It is somewhat 
surprising that this result (50\% chance of a biospherically important event in a half Gyr) is so similar to the original calculation in \citet{Thorsett95}.
At lower fluence, 
$10 $kJ/m$^2$, these probabilities are  higher than 99.8\% (95\%) for 5 Gyr (0.5 Gyr) and thus nearly certain. 
However, the chances of a truly catastrophic event  with  a fluence of $1000 $kJ/m$^2$, are at most 25\% thus making it unlikely. 
These probabilities are of course much larger (see Table.~\ref{table_results}) if we ignore suppression of GRBs in the Milky Way due to large  metallicity.

sGRBs are weaker and as such, even though their  rate is larger than the  rate of LGRBs (and particularly so in the Milky way, because of the metallicity bias) their life threatening effect is negligible as can be seen from  Table~\ref{table_results}.  As {\it ll}GRBs are even weaker their effect is completely  negligible. For completeness we mention that a giant SGR flare would have to be within $\sim 1-2\,$pc from Earth to produce a $100 $kJ/m$^2$ fluence. This is comparable to the  distances between stars in the solar neighbourhood. Consequently giant SGR flares  are  unlikely to cause any significant damage to life.

\begin{table}
\caption{Probability, in \%, of at least one GRB having occurred in the past time $t$ with enough flux to produce significant life extinction. For LGRB we show without parenthesis the probability when there is a 10\% metallicity bias, in parenthesis when there is none. We consider three cases of the GRB fluence on Earth ($10, 100$ and 1000 kJ/m$^2$).}
\begin{center}
\begin{tabular}{cccc}
\hline 
\hline
	& $ t < 5$ Gyr & $t < 1$ Gyr & $t < 0.5$ Gyr  \\
\hline
  $10 $kJ/m$^2$ & & & \\	
\hline
LGRBs &    99.8 (99.95)  & 98.7 (99.90) & 95 (99.80)  \\
sGRBs &    80  & 37  & 22\\
{\it ll}GRBs &     $< 1$ & $< 1$ & $< 1$    \\

\hline
  $100 $kJ/m$^2$ & & & \\	
\hline
LGRBs &    90 (99.8) & 60 (96) & 50 (90)  \\
sGRBs &    14  & 3  & 2\\
{\it ll}GRBs &     $< 1$ & $< 1$ & $< 1$    \\
\hline
  $1000 $kJ/m$^2$ & & & \\	
\hline
LGRBs&    25 (80) & 7 (40) & 4 (25)  \\
sGRBs &    $10^{-2}$  & $2 \times 10^{-3}$  & $10^{-3}$\\
{\it ll}GRBs &     $0$ & $0$ & $0$    \\

\hline
\end{tabular}
\end{center}
\label{table_results}
\end{table}

\section{GRBs and Life in the Galaxy}

We turn now to explore the possible threat caused by GRBs to life elsewhere in the Milky Way, turning to the whole Universe in the next section. Clearly to do so one must assume the lethal radiation dose that will be threatening to life elsewhere. While life can take numerous other forms and could  be much more resilient to radiation than on earth, we make here the conservative assumption that life is rather similar to the one on Earth. This  common assumption  is the basis for searches of Earth like exoplanets as places that harbour life. Under this assumption, we explore what is the likelihood that a nearby GRB results in a dose of 100 as well as 10 and 1000 kJ/m$^2$ in various regions of the Milky Way. 

The stellar density  is significantly larger towards the center of the Galaxy and hence 
the threat to life on most exoplanets, that reside in this region, is much larger. 
Fig. \ref{fig:MWtot} depicts the probability of having one life threatening event within the last\footnote{We use 1 Gyr as a round number to estimate life extinctions that could have cause a massive extinction that terminated life and thus made it unlikely that we find signs of life today.} 1 Gyr as a function of the distance $r$ of an exoplanet from the Galactic center. Also shown is the fraction of the stellar population of the Milky Way within this radius.  A lethal GRBs of $ 100 $kJ/m$^2$ would be more likely than 95\% up to a distance of 2  kpc from the Galactic center in which 25\% of the MW stars reside. 
When considering ${\cal F } = 10$ and $1000$kJ/m$^2$ we find 12  and 0.5 kpc respectively. 
 In agreement with the specific estimates for Earth, events around the Solar distance from the Galactic could be significant but rare and only 
 at a distance $> 10$kpc the threat from GRBs becomes small. 
 Therefore, life can  be preserved with certainty only  in the outskirts of our Galaxy.  In total 90, 40 and 5\% of the exoplanets in the MW
would be exposed to a fluence of 10, 100, and 1000 kJ/m2 from GRBs within a period of 1 Gyr.

Finally, given the LGRBs luminosity function there are practically no lethal events with a distance larger than 30kpc. This implies that nearby small satellite galaxies with a large SFR, like the LMC, are too far to influence life in the  Milky Way.   The fact that the local group is such a low density region containing only two large galaxies (Andromeda and the Milky Way) and with the nearest cluster of galaxies, Virgo,  at $16$ Mpc, i.e. much farther away than the typical inter-galactic distance of 1 Mpc, seems to provide the required  environment to preserve life on Earth. There is no threat from nearby extragalactic bursts.

\section{GRBs and Life in the Universe}

Before concluding we turn now to consider the conditions elsewhere in the Universe. We already mentioned that the local neighbourhood of the Milky Way has a lower density of star forming dwarf galaxies making the Milky Way a more friendly neighbourhood for life.  We can take our calculation one step further and compute  the effective volume in the Universe protected from GRB explosions for life proliferation. This happens for galaxies that produce enough metals  so that their metallicity is at least $1/3$  solar  and their stellar disks are larger than 4 kpc. Using the mass-metallicity relation in  \citet[][their Fig.~6]{panter}   such galaxies must  have stellar masses larger than $10^{10} M_{\odot}$.  This corresponds to a co-moving abundance of $10^{-3}$ galaxies per Mpc$^3$ (see Fig.~3 of \citet[][]{moped}).  This is a factor 10 less than the abundance of most common galaxies. Galaxies friendly to harbor and preserve life will preferably inhabit low density regions in voids and filaments of the cosmic web.

Turning to earlier epochs we may wonder whether life could have existed in the earlier universe? We recall  that the age of the Universe at z=1 is about 6 Gyr so in principle there was enough time for life to evolve even before this redshift; here we note that  the LGRB rate is significantly larger in the past making the GRB threat much more significant. 
Furthermore,  galaxies at high-z are smaller than current ones by a factor of $2-4$ in radius and as such have less room for isolated safe regions like the outskirts of the Milky Way. We conclude that it is impossible to harbor life at $z > 0.5$ as LGRBs will always be 
sufficiently nearby  to life-harboring planets and thus cause life extinctions. It seems the survival of life, as we know it on Earth, was only a recent phenomenon in the history of the Universe caused by the growth of large galaxies. Life forms that might have existed earlier or that exist today in other regions of the Universe that are much more susceptible to significant GRB bombardment  must have been much more resilient to radiation than life on Earth. Of course we do not know whether destruction of a large fraction of life and life forms on a given planet is good or bad for the long-term evolution of higher life-forms on that planet, only that it would be highly damaging for the existing higher life forms, including humans, on our own planet right now, and this is what this study in essence concerns. 

\section{Conclusions}

We have used the latest determination of GRB rates and  luminosities  to estimate the likelihood of them being the source of life extinction on Earth. 
Using also the latest determinations of metallicity of stars in the Milky Way and those of LGRB hosts, we concluded that the likelihood of a GRB producing life extinction on Earth is high. Taking the same lethal dose for extraterrestrial life as for life on Earth
 we have found that GRBs and in particular LGRBs are life threatening in a large part of the Milky Way as well as in many other locations in the Universe. The safest environments to preserve life are the outskirts of large galaxies in low density regions (so that these galaxies don't have ``dangerous"  low metallicity dwarf satellites). It is curious to point out that a cosmological constant of about the same order of magnitud as the present value is essential for the Universe to grow large galaxies and also preserve low density regions at late times $z < 0.5$; the expansion history of a LCDM universe is modified in such a way that it provides enough time at high-z for large under densities and galaxies to grow large. It is also worth mentioning that the damaging nature of GRBs could help explain Fermi's paradox. We will investigate both of these question in detail in a  forthcoming publication.

TP thanks the Institut Lagrange de Paris for hospitality while this work was being completed. This research was supported by the ERC grant GRBs, by the ISF I-Core center of excellence and by an Israel-China grant.  RJ thanks the Royal Society and the ICIC at Imperial College for financial support and hospitality while this work was being completed. We thank Chris Flynn and Luca Casagrande for discussions on the age-metallicity relation of stars in the Milky Way and the anonymous referees for their constructive and useful comments. 

\newcommand{\jcap}{Journal Cosmology and Astroparticle Physics} 
\newcommand{\apjl}{The Astrophysical Journal Letters}
\newcommand{\mnras}{MNRAS}
\newcommand{\aap}{Astronomy \& Astrophysics}
\newcommand{\araa}{ARA\&A}
\newcommand{\pasp}{PASP}

\end{document}